\documentclass[conference,10pt]{article}
\usepackage{url}
\usepackage[backref=page]{hyperref}
\usepackage{cite}
\usepackage{amsmath,amssymb,amsfonts}
\usepackage{algorithmic}
\usepackage{graphicx}
\usepackage{textcomp}
\usepackage{xcolor}
\usepackage[capitalise]{cleveref}
\usepackage{listings}
\usepackage{array}
\usepackage{todonotes}
\usepackage{enumitem}
\usepackage{environ}
\usepackage{caption,subcaption}
\usepackage{authblk}

\crefname{figure}{Fig.}{Figs.}
\crefname{tabular}{Tab.}{Tabs.}
\crefname{section}{Sec.}{Secs.}

\newcolumntype{L}{>{$}l<{$}} 

\def\BibTeX{{\rm B\kern-.05em{\sc i\kern-.025em b}\kern-.08em
    T\kern-.1667em\lower.7ex\hbox{E}\kern-.125emX}}
\begin{document}

\title{Lightning: Multi-GPU Computing using \mbox{Data Annotations}}
\title{Lightning: Scaling the GPU Programming Model for Distributed Multi-GPU Applications}
\title{Lightning: Scaling the GPU Programming Model Beyond a Single GPU}

\author[1,2]{Stijn Heldens\thanks{Corresponding author: s.heldens@esciencecenter.nl}}
\author[2,3]{Pieter Hijma}
\author[1]{Ben van Werkhoven}
\author[1]{\mbox{Jason Maassen}}
\author[1,2]{Rob V.~van Nieuwpoort}

\affil[1]{Netherlands eScience Center}
\affil[2]{University of Amsterdam}
\affil[3]{Vrije Universiteit Amsterdam}

\NewEnviron{smallequation}{%
    \begin{align*}
    \BODY
    \end{align*}
}
\maketitle

\begin{abstract}

The GPU programming model is primarily aimed at the development of applications that run one GPU.
However, this limits the scalability of GPU code to the capabilities of a single GPU in terms of compute power and memory capacity.
To scale GPU applications further, a great engineering effort is typically required:
work and data must be divided over multiple GPUs by hand, possibly in multiple nodes, and data must be manually spilled from GPU memory to higher-level memories.

We present \emph{Lightning}: a framework that follows the common GPU programming paradigm but enables scaling to large problems with ease.
Lightning supports multi-GPU execution of GPU kernels, even across multiple nodes, and seamlessly spills data to higher-level memories (main memory and disk).
Existing CUDA kernels can easily be adapted for use in Lightning, with data access annotations on these kernels allowing Lightning to infer their data requirements and the dependencies between subsequent kernel launches.
Lightning efficiently distributes the work/data across GPUs and maximizes efficiency by overlapping scheduling, data movement, and kernel execution when possible.

We present the design and implementation of Lightning, as well as experimental results on up to 32 GPUs for eight benchmarks and one real-world application.
Evaluation shows excellent performance and scalability, such as a speedup of 57.2$\times$ over the CPU using Lighting with 16 GPUs over 4 nodes and 80 GB of data, far beyond the memory capacity of one GPU.

\end{abstract}


\newcommand{\correction}[1]{#1}

\newpage
\section{Introduction}
\label{sec:introduction}
Many applications in industry/science are nowadays accelerated by \emph{Graphics Processing Units} (GPUs)~\cite{che2009rodinia,danalis2010scalable,vannieuwpoort2011correlatinga,abadi2016tensorflow} and GPUs will likely be used in future exascale systems~\cite{heldens2020landscape}.
A GPU application consists of GPU-specific functions (called \emph{kernels}) that are executed on the GPU by a large number of threads in parallel.
This massive parallelism provides excellent speedups over the CPU, but a single GPU is limited for large problems that exceed the GPU capabilities

There are three orthogonal solutions to increase scalability:
1) spill data from GPU memory to host memory (or even disk),
2) use multiple GPUs within one node,
or 3) use a cluster of GPU-accelerated nodes.
For all these solutions, the programmer must manually split the data into smaller pieces and either stream these pieces through GPU memory, when using a single GPU, and/or distribute them among different memories, when using multiple GPUs.
Data must be communicated between GPUs to maintain data consistency and this intra- and inter-node communication should be overlapped with kernel execution to avoid idle time~\cite{vanwerkhoven2014performance}.
Each kernel launch must also be split into smaller launches and scheduled onto the available GPUs while maintaining correctness.
Additionally, GPU kernel code must be heavily modified to change indexing into data structures and account for offsets in thread indices.
Finally, different tools and libraries must be combined (e.g., MPI, threading, serialization, scheduling, etc.). 
All of this together is a massive engineering effort that leads to complex code that is difficult to develop and maintain~\cite{vanwerkhoven2020lessons}.

Several frameworks have been proposed to aid the development of distributed multi-GPU applications 
either by facilitating local access to remote GPUs for CUDA~\cite{duato2010rcuda, liang2011gridcuda, 
giunta2010gpgpu,oikawa2012dscuda,kim2012snucl,kegel2012dopencl} or OpenCL~\cite{kim2012snucl,kim2016distributed,kegel2012dopencl,alves2013clopencl,grasso2013libwater,aoki2010hybrid,nozal2020enginecl}, 
by abstracting multiple (remote) GPUs into a single virtual device~\cite{kim2011achieving,diop2013distcl,ben-nun2015memory}, 
or by offering special distributed data 
structures~\cite{daskcuda,zhu2014gpuinhadoop,yuan2016sparkgpu}. 
However, no framework alleviate the programmer of all of the above complexities.

In this work, we present \emph{Lightning}: a framework that enables programmers to use a GPU-accelerated cluster in a way that is similar to programming a single GPU, without worrying about low-level details such as network communication, memory capacity, and data transfers.
Lightning supports \emph{distributed kernel launches}, which enable multi-GPU execution of a single kernel,
and \emph{distributed arrays}, which distribute data using a user-specified policy.
Existing CUDA kernels can be used in Lightning with only minor modifications. 
Data access annotations on kernels allow Lightning's runtime system to automatically infer their data requirements, as well as the data dependencies between subsequent kernel launches.
This enable multi-kernel workflows and complex pipelines.

All in all, Lighting provides many features that alleviate programmers from the concerns of multi-GPU programming:

\begin{itemize}[itemsep=.05cm]
\item
Support for \emph{distributed kernel launches} that automatically distribute the work for a single kernel launch across the available GPUs in a cluster.

\item
Existing CUDA kernel code can be reused by making only slight changes and providing data annotations.

\item
Support for multi-dimensional \emph{distributed arrays} that have their data transparently distributed across the cluster.

\item
Data is automatically spilled to higher-level memory, enabling datasets that do not fit into GPU memory.

\item
Data can be (partially) replicated among multiple GPUs and replications are automatically kept consistent.

\item
Focus on asynchronous processing to enable overlapping of scheduling, data movement, and kernel execution.

\item 
Data dependencies between consecutive kernel launches are automatically detected and tasks are executed in parallel in a sequentially consistent order~\cite{lamport1979how}.

\end{itemize}

In this paper, we present the design and implementation of our framework, as well as experimental results for eight benchmarks and one real-world application.
Evaluation shows many excellent results, such as 57.2$\times$ speedup over the CPU using Lightning with 16 GPUs for a dataset of 80 GB, which is far beyond the memory capacity of a single GPU.



\section{Design}
\label{sec:design}
\begin{figure}
  \centering
  \includegraphics[width=.9\textwidth]{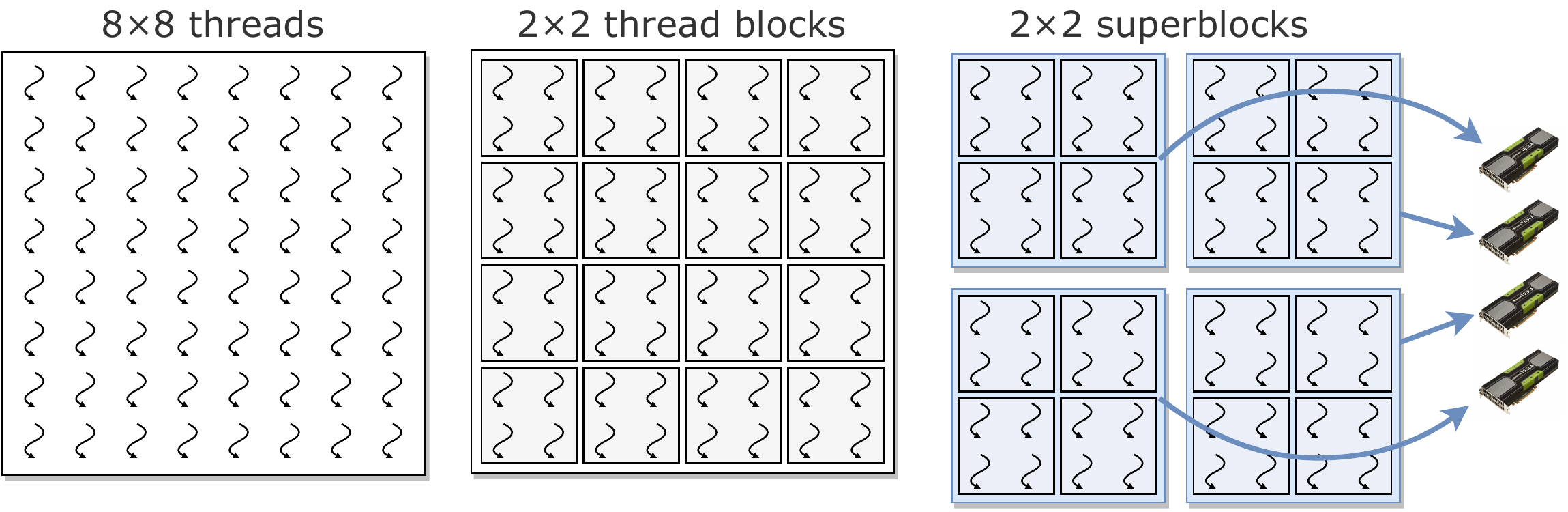}
  \caption{Example of superblock distribution for $8{\times}8$ grid of threads.}
  \label{fig:design_work_distribution}
\end{figure}

In this section, we present the abstractions that Lighting offers to distribute work (\emph{distributed kernel launches}, \cref{sec:design_work}) and distribute data (\emph{distributed arrays}, \cref{sec:design_data}).
These two concepts are united by \emph{data annotations} (\cref{sec:design_annotations}) that allow the \emph{planner} (\cref{sec:design_planner}) to construct an execution plan.

\subsection{Distributed Kernel Launches}
\label{sec:design_work}
In GPU programming (e.g., CUDA or OpenCL), work is performed on the GPU by launching a \emph{kernel} onto the device.
Kernels are GPU-specific functions that are executed by a large number of GPU threads in parallel.
A kernel launch is initiates an $n$-d grid of threads ($n=1,2,3$) where each thread is assigned a unique $n$-d index.
Additionally, the threads are grouped into fixed-sized rectangular \emph{thread blocks}. 
Threads within the same thread block can communicate, while threads from different thread blocks cannot synchronize and run independently of each other\footnote{Recent versions of CUDA added \emph{cooperative} kernels where synchronization across thread blocks is possible, but we focus on conventional kernels.}.

For Lightning, we exploit the fact that thread blocks are independent by distributing the thread blocks of a single kernel launch across multiple GPUs, thus enabling multi-GPU execution of a single kernel.
We call this a \emph{distributed kernel launch}.
The distribution of work is achieved by grouping thread blocks into rectangular disjoint subgrids that we call \emph{superblocks}.
\Cref{fig:design_work_distribution} shows an example.
Each superblock is essentially one job: each is assigned to one GPU in the system and that subset of thread blocks will be executed on that specific GPU.
The superblock distribution must be passed explicitly by the programmer for each kernel launch.

Currently, Lightning supports kernels written in CUDA, although we plan on also supporting other kernel languages.
Small modifications need to be made to the kernel code to make existing CUDA kernels compatible with our framework, such as using Lightning-specific data types (see \cref{sec:design_example})

\begin{figure}
  \centering
  \begin{subfigure}[t]{0.30\textwidth}
    \centering
    \includegraphics[width=.7\textwidth]{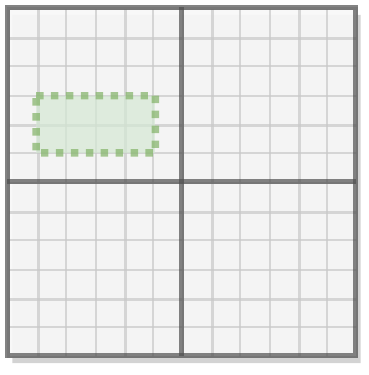}
    \caption{Tile distribution.}
    \label{fig:impl_regions_good}
  \end{subfigure}\hspace{0.01cm}%
\begin{subfigure}[t]{0.30\textwidth}
    \centering
    \includegraphics[width=.7\textwidth]{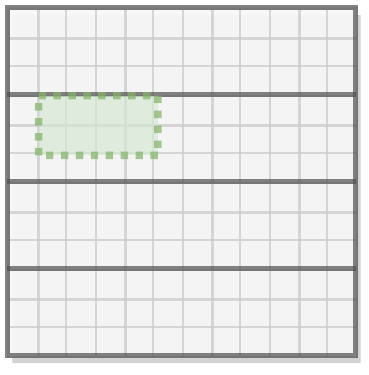}
    \caption{Row-wise dist.}
    \label{fig:impl_regions_good2}
  \end{subfigure}\hspace{0.01cm}%
\begin{subfigure}[t]{0.30\textwidth}
    \centering
    \includegraphics[width=.7\textwidth]{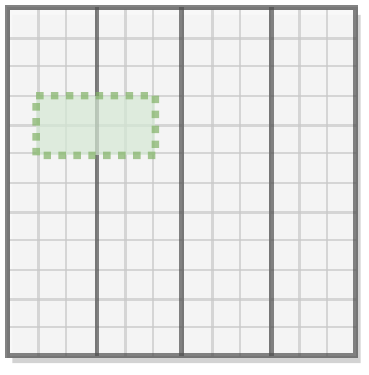}
    \caption{Column-wise dist.}
    \label{fig:impl_regions_bad}
  \end{subfigure}
  \caption{A $12{\times}12$ array partitioned according to three distributions. The black rectangles indicates chunks. The dashed rectangle is an example of the access region of a superblock.}
  \label{fig:impl_regions}
\end{figure}

\subsection{Multi-Dimensional Distributed Arrays}
\label{sec:design_data}
Besides distributing work, it is also necessary to distribute data.
GPU applications typically use multi-dimensional arrays (e.g., vectors, matrices, tensors) as their predominant data structures since they fit the data-parallel model of GPUs.
Therefore, Lighting supports multi-dimensional arrays as its primary data abstraction.
These arrays can be created/deleted dynamically at runtime, have up to three dimensions, and store elements of a primitive type (e.g., \verb=int=, \verb=float=).

Similar to how the threads of a kernel launch must be distributed across GPUs, the data elements of an array also needs to be distributed.
In Lighting, the programmer has to specify the distribution policy for each array.
Such a policy defines a set of rectangular subregions called \emph{chunks} that together cover the entire domain of the array (see \Cref{fig:impl_regions}).
Each chunk is assigned to one GPU in the system.
Several common distributions are included in Lightning (e.g., row/column-wise, tiled) and custom distributions can also be defined.

Whereas superblocks must be disjoint (i.e., each thread is assigned to exactly one superblock), the chunks of one data distribution may overlap (i.e., one data element can be assigned to multiple chunks).
This is useful, for example, for stencil distributions that add a border of halo cells around each tile.
The replicated data elements are automatically kept coherent by Lightning's runtime system.

Although each chunk is assigned to one specific GPU, Lightning will automatically spill the chunk's content from GPU memory to higher-level memories if GPU memory is full (See \cref{sec:implementation_memory}).
It is thus recommended to create chunks with a limited size, allowing the runtime system to overlap kernel execution with transferring chunks into and out of GPU memory.
We found chunks around ${\sim}0.5\text{GB}$ to give good performance (see \cref{sec:evaluation_single}).

\begin{figure}
  \centering
  \includegraphics[width=0.9\textwidth]{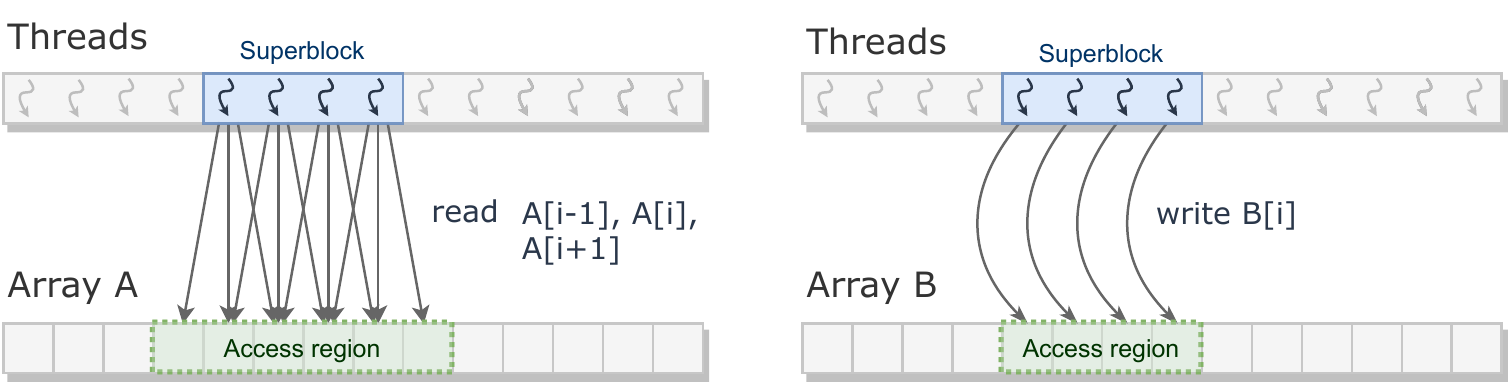}
  \caption{Example of superblock and associated access regions.}
  \label{fig:design_access_pattern}
\end{figure}

\subsection{Data Annotations}
\label{sec:design_annotations}
For each distributed kernel launch, the programmer must specify the arrays that will be accessed by the launched threads.
To be able to distribute these threads across multiple GPUs, we need some way to determine what parts of these arrays are accessed by the threads.
Typically in GPU programming, each thread accesses only a few elements, but this information is normally not encoded into the kernel code.

For Lightning, we define the \emph{access region} of a superblock for an array as the $n$-d dense rectangular area (i.e., lower and upper bounds along each axis) that will be accessed by the threads in that superblock.
As an example, consider a simple 1D stencil kernel where thread $i$ performs $B[i]=A[i-1]+A[i]+A[i+1]$.
\Cref{fig:design_access_pattern} demonstrates the access regions on $A$ and $B$ for one superblock.

To specify these access regions per superblock, Lightning offers a symbolic notation to describe the access pattern of each thread.
By annotating kernels, Lightning can automatically extract the access regions for each superblock.
For example, we can formalize the above stencil access pattern by stating that thread $i$ \verb=read=s elements $A[i-1],{\ldots},A[i+1]$ and \verb=write=s element $B[i]$.
The data annotation in Lightning for this example is as follows:

{\small
\begin{verbatim}
global i => read A[i-1:i+1], write B[i]
\end{verbatim}
}

This annotation should be interpreted as follows.
To the left side of the arrow are variable bindings that, in this case, bind the \verb=global= 1D thread index to variable $i$.
Other possible bindings are \verb=block= (thread block index) and \verb=local= (local index within block).
To the right side of the arrow are statements that describe, for each argument array, the \emph{indices} that are accessed and the \emph{access mode}. 
Each index can either be a single expression or a Fortran-style slice notation ``$\mathit{lower\ bound}:\mathit{upper\ bound}$" (both bounds can be omitted).
Each index expression must be a linear combination of the bound variables to simplify analysis of the access pattern.

For the access mode, there are four supported options:

\begin{itemize}
\item \verb=read=: Access is read-only. Writes are not permitted.
\item \verb=write=: Access is write-only. Reads are not permitted.
\item \verb=readwrite=: Access is both \verb=read= and \verb=write=.
\item \verb=reduce(f)=: Similar to \verb=write=, except `conflicting' writes are reduced (\verb=f= must be \verb=+=, \verb=*=, \verb=min=, or \verb=max=).
\end{itemize}

Another example of an annotation is for a naive matrix multiplication kernel performing $C=AB$ where thread $(i,j)$ writes entry $C_{ij}$, reads row $i$ of $A$, and reads column $j$ of $B$.

{\small
\begin{verbatim}
global [i, j] => read A[i,:], read B[:,j], write C[i,j]
\end{verbatim}
}

Yet another example is a reduction of matrix $A$ along the columns to a vector \verb=sum=.
Thread $(i,j)$ reads $A_{ij}$, threads cooperatively reduce values and write their results to $\texttt{sum}_i$.

{\small
\begin{verbatim}
global [i, j] => read A[i,j], reduce(+) sum[i]
\end{verbatim}
}

For the reductions, Lightning internally allocates temporary memory to which the threads can write their local results.
Afterward, Lightning performs a multi-level reduction. 

\begin{figure}
  \centering
  \includegraphics[width=.9\textwidth]{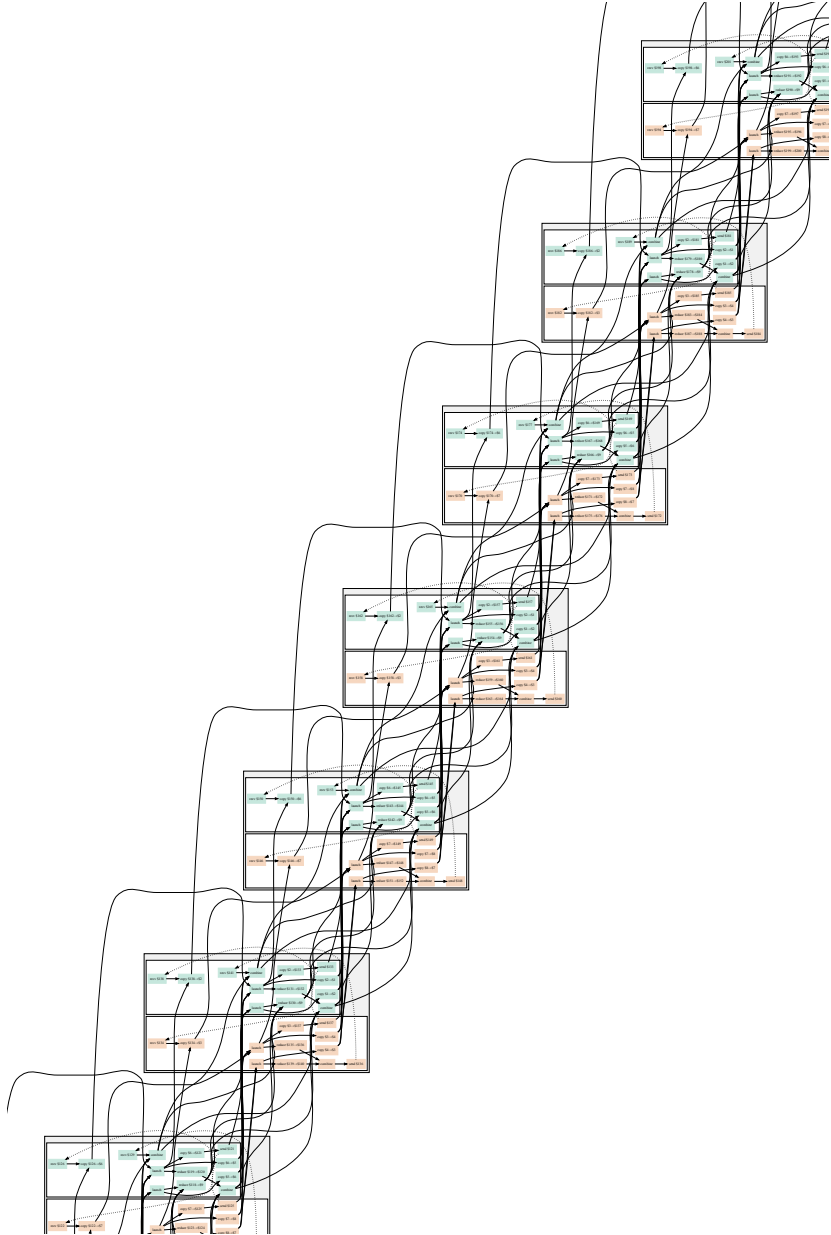}
  \caption{DAG created for stencil kernel (\cref{lst:stencil_host}). Shows four iterations on two workers with two GPUs each. Large boxes represent a distributed kernel launches and smaller boxes represent individual tasks (color indicates worker id).}
  \label{fig:stencil_dag}
\end{figure}

\subsection{Execution Planner}
\label{sec:design_planner}
For each distributed kernel launch, Lightning will construct an \emph{execution plan}.
Such a plan consists of a \emph{directed acyclic graph} (DAG) for each node in the system containing the tasks for that node and the dependencies between tasks.
Examples of DAG tasks are \verb=Execute= a kernel, \verb=Create/Delete= a chunk, \verb=Copy= data between chunks and \verb=Send=/\verb=Recv= chunks between nodes.
\Cref{fig:stencil_dag} shows an example of an execution plan. 

Execution plan construction is performed by the \emph{planner}.
First, the planner divides the kernel launch into superblocks.
For each superblock, the planner processes each argument array.
For each argument, the planner first evaluates the data annotation to determine the access region and then queries the array's data distribution to determine which chunks intersect the access region.
In the common case, data is distributed such that there will be at least one chunk enclosing this access region (see \cref{fig:impl_regions_good,fig:impl_regions_good2}).
If that chunk is assigned to the superblock's GPU, then the chunk can be used directly.
Otherwise, it must be copied between GPUs, or even between nodes, by inserting \verb=Copy=/\verb=Send=/\verb=Recv= tasks into the DAG.
For \verb=write= accesses, the planner also inserts proper data transfers to update replicated data elements.

In exceptional cases, the access region might intersect with multiple chunks (\cref{fig:impl_regions_bad}).
For \verb=read= access, the planner assembles a temporary chunk from the contents of the intersected chunks.
For \verb=write= access, the planner creates a temporary uninitialized chunk and afterward scatters its content. 
While this procedure might be inefficient, it means data distributions only affect the \emph{performance} of an application and not the \emph{correctness}.
This provides separation of concerns: programmers can first develop the application and later tune the work/data distributions to maximize performance.

The planner handles \verb=reduce= accesses separately.
For each superblock, a temporary chunk is created to hold the block-level partial results.
Afterward, the planner inserts reduction tasks to hierarchically reduce the partial results: 
first the results for one superblock, then for one GPU, then for each node, and finally reducing the results across all nodes.

After the execution plan for one distributed kernel launch has been constructed, the DAGs are immediately submitted to the nodes in the system.
Execution on the host continues, allowing additional kernels to be launched.
This increases efficiency since it overlaps plan construction with kernel execution and data movement on the nodes.
However, this means that tasks from the previously submitted DAGs might not have finished when the next kernel is already being planned.
To solve this problem, the planner analyzes the dependencies between consecutive distributed kernel launches and inserts dependencies from previously submitted tasks when there are data conflicts on chunks (i.e., read-write/write-write/write-read conflicts).
Essentially, the planner incrementally builds a large DAG from many smaller DAGs. 

\subsection{Scope and Limitations}
While the design of Lightning is versatile, there are limitations.
First, while the data annotations are simple and expressive, they require the data access pattern to be predictable and derivable from the thread/block indices.
This is the case for regular algorithms, for example those from linear algebra. 
However, data-dependent problems generally cannot be expressed. 
In some cases, they can be described imprecisely, resulting in a performance penalty.
For example, sparse matrix-vector multiplication (SpMV) performs unstructured reads on the input vector, but can be still be expressed by overestimating the access region to be the entire vector (see \Cref{sec:evaluation_benchmarks}).

Second, Lightning supports multi-dimensional arrays since they fit well into the data-parallel model of GPUs.
Irregular data structures, such as linked lists or graphs, are unsupported.
Additionally, the data access patterns on the arrays must be dense and rectangular, other patterns cannot be expressed for now (e.g., triangular, diagonals, indirection).

Third, users must manually annotate their code.
Automatic extract of data annotations using static code analysis, while interesting, is out of scope for this manuscript.

\section{Implementation}
\label{sec:implementation}

\begin{figure}
  \centering
  \includegraphics[width=\textwidth]{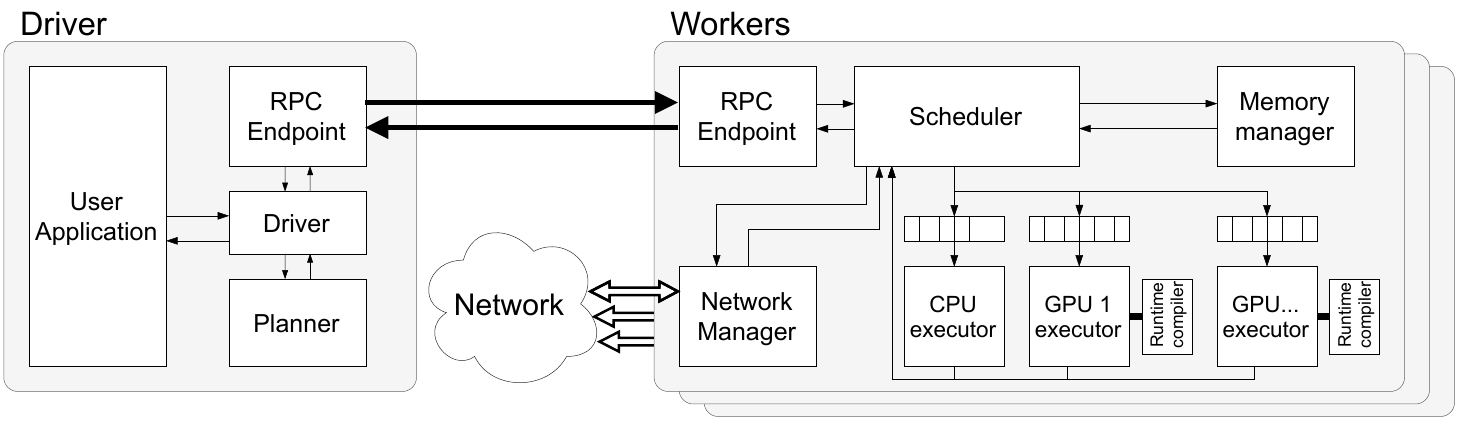}
  \caption{Overview of Lightning's runtime system.}
  \label{fig:impl_architecture}
\end{figure}

In this section, we discuss the implementation of Lightning.

\subsection{Overview}
\Cref{fig:impl_architecture} shows the software architecture of Lightning.
Our system runs on a cluster of \emph{worker nodes} which are managed by one central \emph{driver program}.
The driver acts as the centralized component in the system: it coordinates the workers, maintains bookkeeping of the distributed arrays, and builds the execution plans.
Each worker node is equipped with one or more GPUs and the workers execute the commands submitted by the driver.
In our implementation, the driver program also runs on the first worker node, meaning there is no network overhead when using just a single node.

The driver also runs the user's application and each call made by the application into the system is handled by the driver.
For instance, when the application creates an array, the driver maintains the associated metadata and requests the workers to allocate the chunks in memory.
When the application launches a kernel, the driver builds the execution plan and submits the resulting DAGs to the workers.
The rationale behind this choice is that it matches the conventional model of GPU programming where a central host offloads compute-intensive tasks to a discrete GPU.

\subsection{Communication \& Data Movement}
Lightning uses MPI for the network layer.
We use a simple RPC protocol on top of MPI for the communication between the driver and the workers, since this traffic consists solely of small control messages.
For communication between the workers themselves, we use non-blocking MPI point-to-point primitives, since this network traffic consists of bulk data exchanges.
We assume workers are connected through a fast interconnect (e.g., InfiniBand), although any MPI implementation is compatible with Lightning.

Data transfers between host memory and GPU memory are performed using asynchronous memory copies to allow overlapping data movement with kernel execution.
Data transfers between two GPUs on the same node are performed using asynchronous peer-to-peer copies, which uses DMA (\emph{direct memory access}) to directly copy between GPUs.
For transfers between GPUs on different nodes, data is staged in host memory and transferred using MPI.
This gives sufficient performance since Lighting effectively exploits asynchronous processing to overlap data movement with kernel execution.

\subsection{Scheduling}
\correction{
For each distributed kernel launch submitted by the application, the driver requests the execution planner to construct an execution plan, consisting of a DAG of tasks for each worker.
The driver submits these DAGs to the workers and each worker has its own \emph{scheduler} to schedules these tasks onto the local resources.
The actual \emph{scheduling} of the DAG is thus done by the workers themselves, while the driver only \emph{plans} the DAG.
This is important since DAG tasks can be small (in the order of milliseconds) and centralized scheduling would quickly become a bottleneck.
}

Initially, each task must wait until its predecessor tasks finish.
Once a task's dependencies (i.e., predecessor tasks) complete, the task is ready to be executed.
First, the {scheduler} submits the task to the \emph{memory manager}  for \emph{staging}.
Each task is associated with several chunks it will access and staging entails that these chunks must be materialized in the requested memory space (see \Cref{sec:implementation_memory}).
Next, after staging completes, the \emph{scheduler} queues the task at the appropriate \emph{executor} (i.e., CPU, GPU, or network).
Finally, once the task finishes execution, the scheduler requests the memory manager to \emph{unstage} the tasks (i.e., release the task's chunks) and checks which successor tasks are not ready for execution.

When multiple tasks become ready simultaneously, the scheduler selects one arbitrary task without further considerations.
We found that this performs adequately in practice since Lightning effectively exploits asynchronous processing. 
For future work, we will explore more complex scheduling policies that consider, for example, data locality  or task priority.

\subsection{Memory Management}
\label{sec:implementation_memory}
Every worker has its own \emph{memory manager} that maintains the bookkeeping of all local chunks and where they are allocated. 
Each chunk can be allocated in host memory, GPU memory, or disk.
The memory manager automatically moves chunks between these different memory spaces when required.

For each task that gets staged, the memory manager's responsibility is to materialize the chunks associated with the task.
First, memory must be allocated for chunks that are currently not allocated in the requested memory space.
The memory manager uses pre-allocated memory pools because we found allocations of device memory and page-locked host memory to be expensive.
It is important that all the task's chunks are allocated in one action to prevent deadlocks.
If memory is full, previously allocated unused chunks are evicted in least-recently used fashion to higher-level memory (i.e., GPU to RAM, RAM to disk).

Second, if a the data in the allocated chunk was previously evicted, data must be copied back from the higher-level memory.
All data transfers performed by the memory manager are asynchronous.
It is important that a sufficient number of tasks is being staged concurrently to enable overlapping work performed by the executors with the staging of future work by the memory manager.

The scheduler must throttle the number of tasks that are staged simultaneously at any moment in time since this number presents a trade-off.
On the one hand, allowing too few concurrently staged tasks prohibits overlapping data transfers with task execution.
However, on the other hand, allowing too many leads to contention where tasks are staged too far ahead of time.
Our current implementation uses a simple heuristic to throttle the number of concurrently staged tasks: 
the total memory footprint of tasks that are staged onto one resource simultaneously cannot exceed some predefined threshold.
We found a threshold of 2 GB to work well in practice.

\begin{figure*}
\centering
\begin{lstlisting}
__global__ void stencil(

        int n,
        float *output,
        const float *input
) {
  int i = blockDim.x  * blockIdx.x + threadIdx.x;
  if (i >= n) return;

  float left = i-1 >= 0 ? input[i-1] : 0;
  float mid = input[i];
  float right = i+1 < n ? input[i+1] : 0;
  float new_val = (left + mid + right) / 3.0;

  output[i] = new_val;
}
\end{lstlisting}
\caption{Original CUDA source code.}
\label{lst:example_kernel_original}
\quad
\begin{lstlisting}
<@\color{red}{\texttt{\_\_device\_\_}}@> void stencil(
        <@\color{red}{\texttt{\textbf{dim3} virtBlockIdx,}}@>
        int n,
        <@\color{red}{\texttt{\textbf{lightning::Vector<float>}}}@> output,
        <@\color{red}{\texttt{\textbf{const lightning::Vector<float>}}}@> input
) {
  int i = blockDim.x  * <@\color{red}{\texttt{virtBlockIdx}}@>.x + threadIdx.x;
  if (i >= n) return;

  float left = i-1 >= 0 ? input[i-1] : 0;
  float mid = input[i];
  float right = i+1 < n ? input[i+1] : 0;
  float new_val = (left + mid + right) / 3.0;

  output[i] = new_val;
}
\end{lstlisting}
\caption{Modified code from \cref{lst:example_kernel_original} (Changes in red).}
\label{lst:example_kernel_modified}
\begin{lstlisting}[language=C]
extern "C" __global__ void stencil_wrapper_ftpyotpf8VofcBIdGGEfxrlOdmfpzbWY(
  int32_t n, 
  float *const output_ptr, 
  const float *const input_ptr
) {
  // Worker-specific constants
  const uint32_t block_offset_x = 1024, block_offset_y = 0, block_offset_z = 0;
  const size_t input_offset_0 = 1023, input_strides_0 = 1;
  const size_t output_offset_0 = 1024, output_strides_0 = 1;

  // Prepare arguments
  dim3 virtual_block_index(block_offset_x + blockIdx.x, 
    block_offset_y + blockIdx.y, block_offset_z + blockIdx.z);
  ::lightning::Array<float, 1> output(
    output_ptr - output_offset_0 * output_strides_0, {output_strides_0});
  const ::lightning::Array<float, 1> input(
    input_ptr - input_offset_0 * input_strides_0, {input_strides_0});

  // Call user kernel
  stencil(virtual_block_index, n, max_diff, output, input);
}
\end{lstlisting}
\caption{Example of the generated wrapper kernel used internally by Lightning at runtime for \Cref{lst:example_kernel_modified}.}
\label{lst:example_kernel_wrapper}
\end{figure*}

\subsection{Runtime Kernel Compilation}
Lightning supports existing GPU kernels written in CUDA, with minor modifications.
To illustrate these changes that one must make, we use an example of a simple stencil operation (see \Cref{lst:example_kernel_original}).
Three changes must be made by the user (see \cref{lst:example_kernel_modified}) before this kernel can be used within Lightning: 

\begin{itemize}
\item
Change the declaration from \verb|__global__| (kernel function) to \verb|__device__| (device function). 

\item
Explicitly take the block index as a parameter.
This is required since this index will be virtualized, so the physical block index (\verb=blockIdx= in CUDA) is incorrect.

\item
Change arguments from raw data pointers to Lightning-specific data types (\verb=Scalar=, \verb=Vector=, \verb=Matrix=, \verb=Tensor= for $0, 1, 2$, or $3$-D arrays).
These types overload several operators and can be accessed like regular arrays without changing their indexing.

\end{itemize}

The system performs runtime compilation, meaning that the source code of a CUDA kernel must be provided at runtime; each worker in the system compiles \emph{a local version} of the code and loads the resulting kernel into the GPU at runtime.
The regular NVIDIA  CUDA compiler is used at runtime for compilation. 
The advantage of runtime  over ahead-of-time compilation is that any runtime constant (for index calculations) can be inserted into the kernel code at compile-time to minimize the overhead of our framework.
In \Cref{sec:evaluation_app}, we show that runtime compilation means the overhead of Lightning over directly using CUDA is small.

The user's kernel is not called directly, but instead, Lightning generates a wrapper kernel that performs some steps before calling the user's kernel.
First, an offset is added to the physical block index and the user's kernel is called with this virtual block index as its first argument.
This solves the problem that CUDA always numbers thread blocks from zero.
Second, the wrapper is passed chunks that correspond to subregions of larger arrays and, 
to give the illusion that the full array can be indexed, offsets must be subtracted from the \emph{global} array indices to obtain the \emph{local} chunk indices.
To solve this, Lightning uses special data types that subtract an offset from the chunk's memory address.
These data types subtract this offset once on construction, 
meaning that there is no performance cost on element access.

\Cref{lst:example_kernel_wrapper} shows the wrapper kernel generated by Lightning internally at runtime for \Cref{lst:example_kernel_modified}.
This code is shown here for academic purposes, it is not intended to be seen by the end-user.
Lines 7-9 show generated constants that are specific for one worker.
Lines 12-14 show how the virtual block indices (add offsets) and data types (subtracts offsets) are constructed.
Line 17 calls the user's kernel with the correct arguments.

\begin{figure}
\centering
\begin{lstlisting}[language=Caml]
let stencil = CudaKernelDef::from_file("stencil.cu")
  .param_value("n", DTYPE_INT)
  .param_array("output", DTYPE_FLOAT)
  .param_array("input", DTYPE_FLOAT)
  .annotate("global i => read input[i-1:i+1],
                         write output[i]")
  .compile(context)?;

let devices = context.system().devices();
let n = 1_000_000;
let data_dist = StencilDist::new(64_000, 1, devices);
let input = context.ones(n, data_dist)?;
let output = context.zeros(n, data_dist)?;

let work_dist = BlockDist::new(64_000, devices);
for _ in 0..10 {
 stencil.launch(n, 16, work_dist, (n, output, input))?;
 swap(input, output);
}

context.synchronize()?;
\end{lstlisting}
\caption{Host code sample for the stencil kernel (\Cref{lst:example_kernel_modified}).}
\label{lst:stencil_host}
\end{figure}

\subsection{Host Code Sample}
\label{sec:design_example}

\Cref{lst:stencil_host} shows an example of the host application for the stencil kernel from \Cref{lst:example_kernel_modified}.
Lightning's runtime system is implemented in the Rust programming language.
For now, host code also needs to be Rust, but library bindings for other programming languages are part of future work.

First, the kernel source code must be loaded for runtime compilation.
Line~1 loads the CUDA kernel code from a separate file \texttt{stencil.cu} (shown in \Cref{lst:example_kernel_modified}),
lines~2-6 provide the definition of the kernel's signature (i.e., parameters and data annotations),
and line~7 submits the kernel code to the workers for compilation. 

Next, the data distributions and arrays must be defined.
Line~11 defines the data distribution to be used: a stencil distribution with a chunk size of $64\,000$ ($256$kB) distributed round-robin across all GPUs.
Lines~12-13 define two vectors of size \texttt{n} having the above data distribution.

Finally, distributed kernels launches can be submitted.
Line~16 defines the superblock distribution to be used: a block distribution having $64\,000$ threads per superblock.
Line~17 launches the stencil kernel 10 times with the provided superblock distribution.
Kernel launches are asynchronous to the driver, so line~21 blocks the driver until work completes.


\section{Experimental Evaluation}
\label{sec:evaluation}
In this section, we present performance results for Lightning.
\Cref{sec:evaluation_setup} describes the experimental setup.
\Cref{sec:evaluation_benchmarks,sec:evaluation_single,sec:evaluation_multi,sec:evaluation_dist} present eight benchmarks on three platforms: one node with one GPU, one node with 4 GPUs, and a cluster with 32 GPUs.
\Cref{sec:evaluation_app} presents a full application for geospatial cluster analysis that was ported to Lightning

\subsection{Experimental Setup}
\label{sec:evaluation_setup}
We performed experiments at Microsoft Azure US East on nodes of type \verb=NC24rsV2=.
Each node contains an Intel E5-2690 CPU with 24 cores, 448 GB of memory, 3TB of temporary SSD storage, and 4 NVIDIA Tesla P100 GPUs with 16 GB memory each.
The GPUs likely utilize PCIe 3.0 x16 (indicated by bandwidth benchmarks~\cite{vanwerkhoven2014performance}) and nodes are connected to each other by InfiniBand FDR, providing high bandwidth.
The software used was Ubuntu 20.04, Rust 1.56, CUDA 11.4, and OpenMPI 4.0.3.

Presented execution times are the average over 5 runs.
One initial untimed run is always performed to warm up the system.
Each run is measured from the moment that the first distributed kernel launch is submitted until the moment that the driver signals the application that all workers finished. 
This timing thus includes the overhead for execution plan construction.
We emphasize that the code is not changed when moving between different platforms.

\subsection{Benchmarks}
\label{sec:evaluation_benchmarks}
To evaluate the performance of Lightning in different scenarios, we selected eight CUDA kernels representing different workloads.
The kernels were taken from various sources and adapted to make them suitable for Lightning (similar to the example in \cref{lst:example_kernel_modified}).
The first four benchmarks are compute-intensive (i.e., high arithmetic intensity), while the latter four are data-intensive.
For each benchmark, we define a parameter $n$ (the \emph{problem size}) such that the amount of \emph{work} scales linearly with $n$. 
However, it is important to note that the amount of \emph{data} need not necessarily scale linearly with $n$.

\begin{itemize}

\item \textbf{MD5} (from SHOC~\cite{danalis2010scalable}) calculates $n$ MD5 hashes in parallel. 
Work is divided into superblocks of 5B threads each.
No data is involved (except one search hash), thus this is a purely compute-oriented benchmark.

\item \textbf{N-Body} (from CUDA samples~\cite{cuda}) performs 10 iterations of an all-pair gravitational simulation.
The benchmark generates $\sqrt{n}$ bodies, so the number of pair-wise interactions (i.e., workload) equals $n$.
The data is replicated (data size is small) and the work is divided equally.

\item \textbf{Correlator} (from van Nieuwpoort et al.~\cite{vannieuwpoort2011correlatinga}) calculates the correlation between each pair of 256 radio antennas for $n$ frequency channels.
The data/work is partitioned with 64 frequency channels per chunk.
Note that the original code used a 2D grid of threads and mapped each 2D thread index to a 3D index.
This access pattern could not be expressed using Lightning's annotations, thus the code was simplified to use a 3D thread grid instead.

\item \textbf{K-Means} (from Rodinia~\cite{che2009rodinia}) is an iterative clustering algorithm commonly used in data mining.
The benchmark uses $n$ records (each having 4 features), finds $k{=}40$ clusters, and performs 5 iterations.
The distribution uses 25M records per chunk.
The original code performed the center calculation on the CPU, but our code utilizes the GPU thanks to Lightning's support for reductions.

\item \textbf{HotSpot} (from Rodinia~\cite{che2009rodinia}) models thermal simulation of an integrated circuit by performing 10 iterations of a $3{\times}3$ stencil.
The benchmark uses a $\sqrt{n} \times \sqrt{n}$ grid (total of $n$ grid points) with a column-wise distribution such that each chunk contains 50M points.
Halo elements are exchanged in each iteration.

\item \textbf{GEMM} (handwritten, based on Volkov et al.~\cite{volkov2008benchmarking}) performs a dense matrix-matrix multiplication $C=AB$.
Matrices $A$, $B$, and $C$ have size ${\sqrt[3]{n}} \times \sqrt[3]{n}$ to ensure the total workload is $n$ (cubic time complexity).
The matrices are partitioned row-wise with 250M elements per chunk.
The work partitioned in the same way, meaning that the data for $A$ and $C$ is available locally, but the entire matrix $B$ must be exchanged between GPUs, making this a very communication-intensive benchmark.

\item \textbf{SpMV} (from SHOC~\cite{danalis2010scalable}) performs repeated multiplication of a sparse $\sqrt{n} \times \sqrt{n}$ matrix with a dense vector of size $\sqrt{n}$.
Ten iterations are performed, where the output of each operation is used as the input for the next iteration.
The vector is broadcast after each iteration.
The matrix is stored in ELL format and its density is $0.1\%$ (i.e., the fraction of non-zeros).
The vectors are replicated while the matrix is row-wise distributed with 100M elements per chunk.

\item \textbf{Black-Scholes} (from CUDA samples~\cite{cuda}) computes call-put prices of $n$ financial options using the Black-Scholes model.
This problem is embarrassingly parallel since $n$ models can be calculated in parallel.
Each chunk contains 100M options.

\end{itemize}

\begin{figure}
\centering
\begin{minipage}[t]{0.45\textwidth}
\centering
\includegraphics[height=.7\textwidth,trim=0 0 465 170,clip]{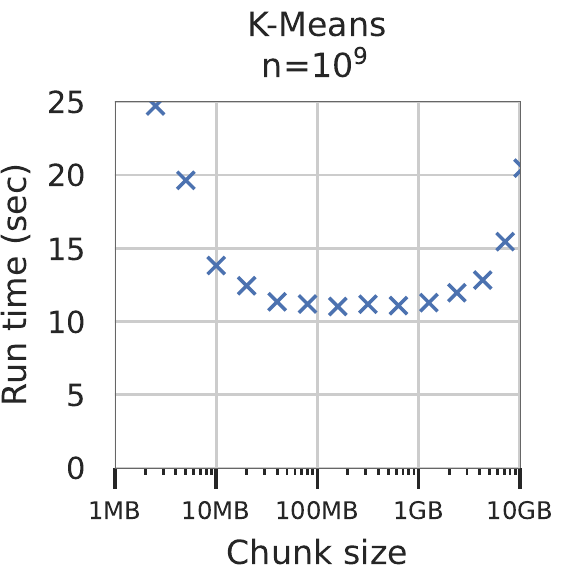}
\caption{Throughput versus chunk size for one GPU. Note the logarithmic x-axis.}
\label{fig:evaluation_chunk_kmeans}
\end{minipage}\hspace{0.03\textwidth}%
\begin{minipage}[t]{0.45\textwidth}
\centering
\includegraphics[height=.7\textwidth,trim=0 0 465 170,clip]{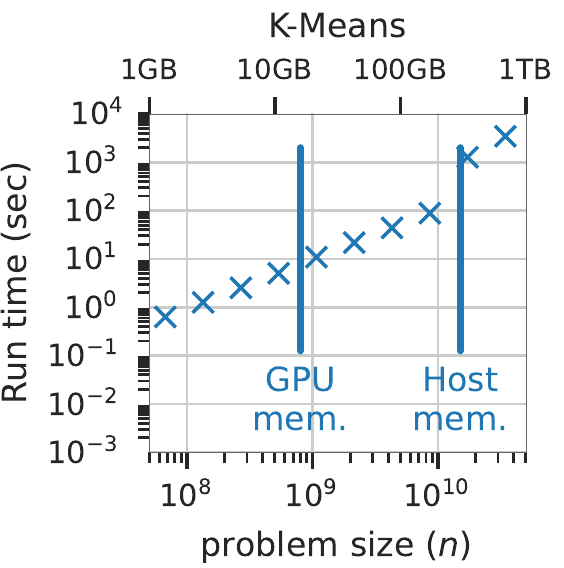}
\caption{Run time versus problem size for one GPU. Note the logarithmic axes.}
\label{fig:evaluation_time_kmeans}
\end{minipage}%
\end{figure}

\subsection{Single GPU}
\label{sec:evaluation_single}
In this section, we present results when using a single GPU.
To understand the sensitivity of performance to the chunk size, we evaluated the \verb=K-means= application for different chunk sizes for a problem size that just exceeds GPU memory ($n{=}{10}^9$).
The results in \Cref{fig:evaluation_chunk_kmeans}  show that the chunk size should not be too small (i.e., ${<}50 MB$, leads to scheduling overhead) or too big (i.e., ${>}5 GB$, prohibits overlapping data transfers and kernel execution).
However, a wide range of chunk sizes gives similar performance indicating that performance is not sensitive to the chunk size.

Next, we consider different problem sizes.
For example, \cref{fig:evaluation_time_kmeans} shows the execution time versus the problem size for \verb=K-Means=.
As anticipated, the run time scales linearly with the problem size $n$.
To ease further analysis, we define \emph{throughput} as the problem size divided by the execution time (i.e., number of items processed per second).
Note that the definition of the \emph{problem size} differs per benchmark, thus throughputs are not comparable across benchmarks.

\begin{figure*}
\centering
\includegraphics[width=\textwidth]{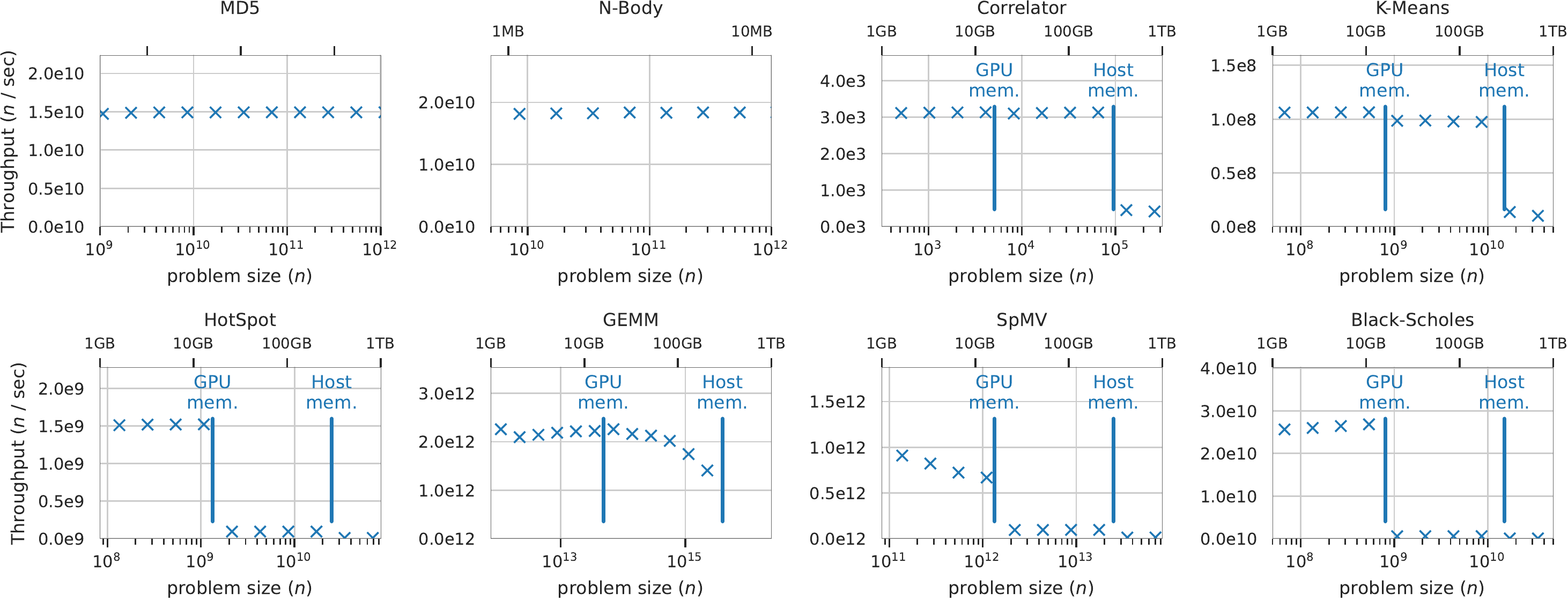}
\caption{Throughput versus problem size when using a single GPU.  Two vertical lines indicate the largest problem that fits into GPU memory and host memory (\texttt{N-Body} and \texttt{MD5} always fit). 
The bottom axis shows problem size ($n$) while the top axis shows the corresponding memory footprint. Note the logarithmic scale on the x-axis.}
\label{fig:evaluation_single_throughput}
\end{figure*}

\Cref{fig:evaluation_single_throughput} shows this throughput metric for different problem sizes for each of the eight benchmarks.
Nearly all benchmarks show that the throughput is roughly consistent across different problem sizes as long as data fits into GPU memory.
This is expected since the workload of each benchmark scales linearly with $n$.
One exception is \verb=SpMV= which performs better for smaller problem sizes.
Further examination revealed that this is due to cache behavior since this benchmark involves random accesses and data fits better into caches for smaller $n$.

Each plot also shows vertical bars indicating the largest problem that fits into GPU memory (first bar) and host memory (second bar).
\verb=MD5= and \verb=N-Body= always fit into GPU memory.
For large problems, Lightning must spill chunks to host memory (or even disk) and transfer them back to GPU memory as needed, which incurs a performance hit.
We see that spilling to disk is never worthwhile due to limited disk bandwidth. 
It is possible that faster non-volatile storage could make this useful over the regular SSDs in this node.

Spilling to host memory, on the other hand, is beneficial for three benchmarks: \verb=Correlator=, \verb=K-means=, \verb=GEMM=.
For these benchmarks, Lightning can overlap kernel execution with the data transfers between GPU and host memory.
For example, for \verb=Correlator=, throughput drops by just $8.8\%$ from $n=16384$ (8.6 GB) to $n=32768$ (17.2 GB).
However, for the three data-intensive benchmarks (\verb=HotSpot=, \verb=SpMV=, \verb=BlackScholes=), overlapping kernel execution with data transfers is not possible since these applications do not perform sufficient work per byte transferred of the PCIe bus.
For example, for \verb=BlackScholes= with $n=0.5 \times 10^9$, the dataset of $10.7$ GB is processed in $20.2$ ms, meaning that PCIe should provide a bandwidth of $530$ GB/s to keep up, over an order of magnitude more than what PCIe 3.0 x16 is capable of.

We conclude spilling to host memory is beneficial for compute-intensive applications.
For data-intensive applications, the PCIe bus provides insufficient bandwidth to overlap data transfers.
We can avoid spilling by using multiple GPUs since this provides more (combined) GPU memory.

\begin{figure*}
\centering
\includegraphics[width=\textwidth]{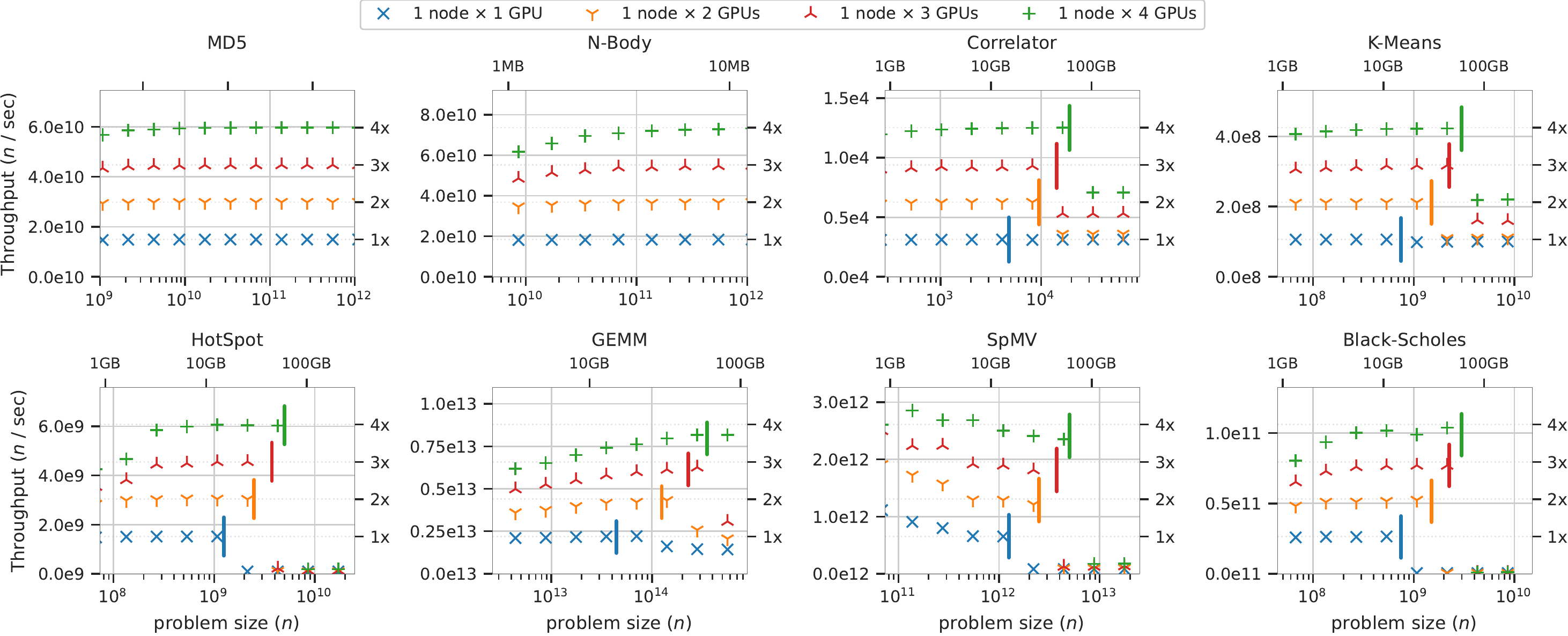}
\caption{\correction{Throughput versus problem size when using a multi-GPU node. The left labels indicate throughput, bottom labels indicate problem size ($n$), top labels indicate memory footprint, right labels indicate speedup over the baseline throughput.}}
\label{fig:evaluation_multi_speedup}
\centering
\includegraphics[width=\textwidth]{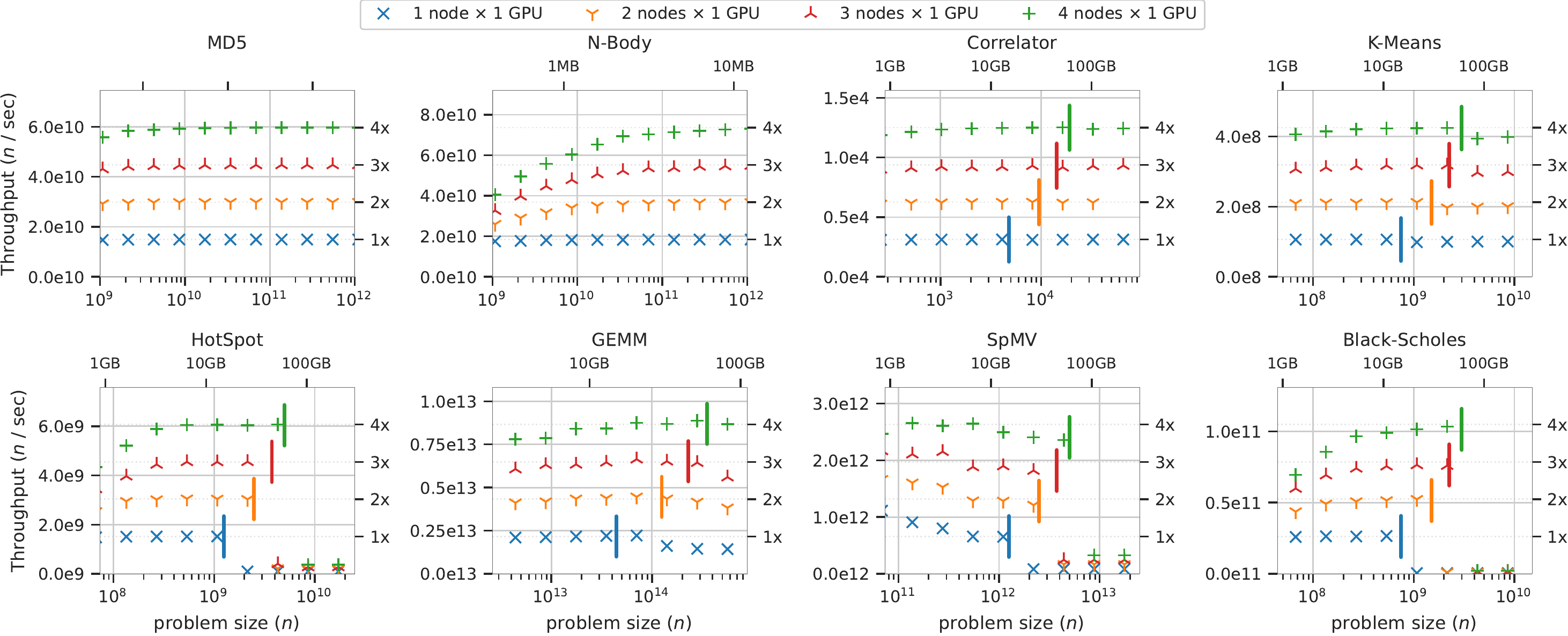}
\caption{\correction{Throughput versus problem size when using multiple nodes (one GPU per node).}}
\label{fig:evaluation_dist_speedup}
\centering
\includegraphics[width=.85\textwidth]{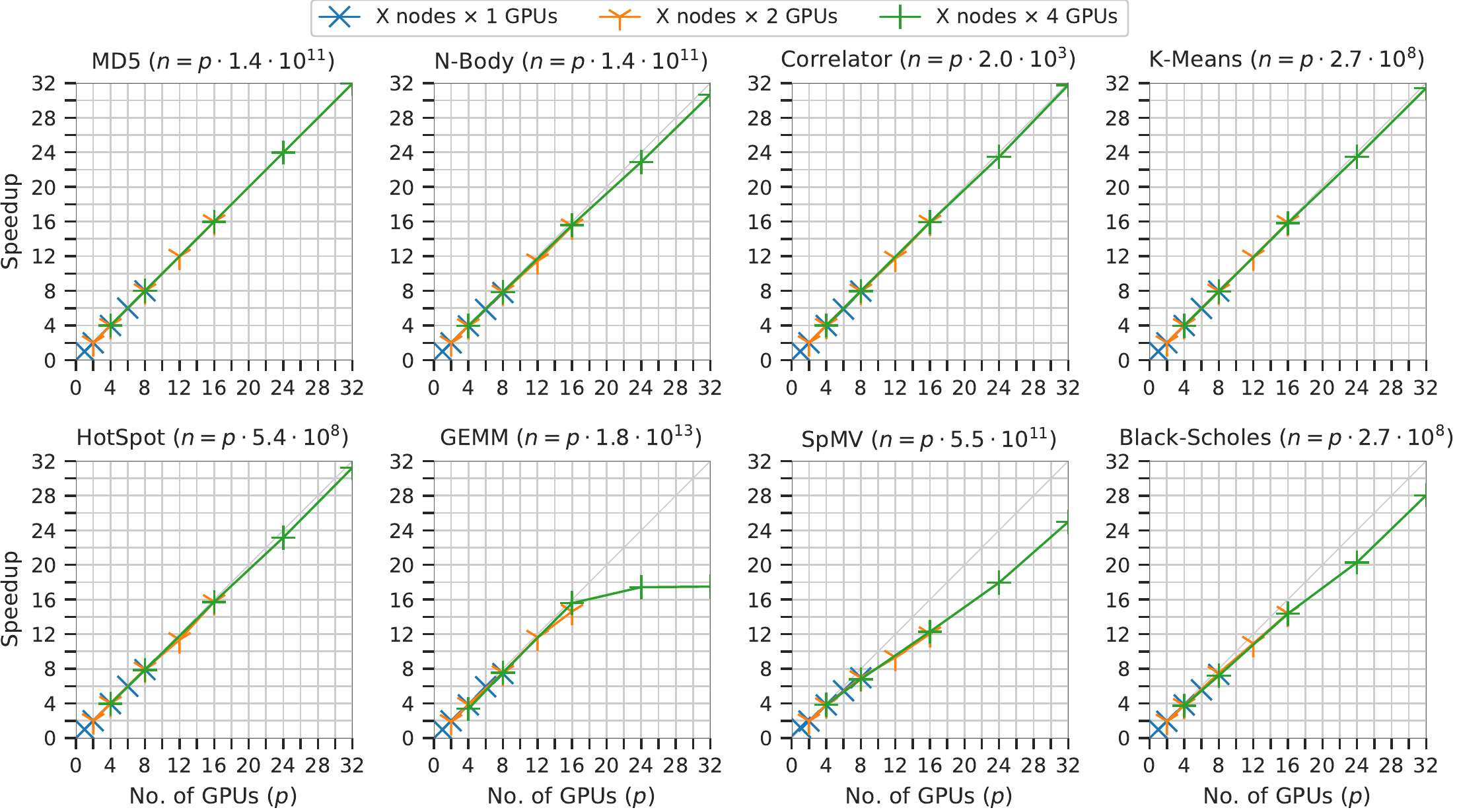}
\caption{Weak scaling experiment. Speedup versus number of GPUs ($p$) for 1, 2, or 4 GPUs per node.}
\label{fig:speedup_small}
\end{figure*}

\subsection{Multiple GPUs} 
\label{sec:evaluation_multi}
Next, we present results when using multiple GPUs on a single node.
\Cref{fig:evaluation_multi_speedup} shows the throughput for up to 4 GPUs for different problem sizes.
Ideally, the throughput should $p$ times higher for $p$ GPUs (i.e., speedup of $p$).
To give an indication of speedup, the labels on the right indicate multiples of the \emph{baseline throughput} (i.e., throughput obtained using one GPU for the largest problem size that still fits into GPU memory).

The plots show that Lightning obtains excellent speedups for all benchmarks.
For example, for \verb=Correlator=, \verb=K-means= and \verb=MD5=, speedups are nearly perfect: these benchmarks are compute-intensive and thus scale well.
For other benchmarks, such as \verb=GEMM= and \verb=N-Body=, speedups are good except for smaller problem sizes.
These benchmarks involve communication, leading to synchronization overhead for small inputs.

Multiple GPUs mean more (combined) GPU memory, indicate in \Cref{fig:evaluation_multi_speedup} by the vertical bars that move further to the right as more GPUs are utilized.
Larger problems can be processed before data is spilled to host memory.
The benchmarks for which spilling was not beneficial in the previous section (\verb=HotSpot=, \verb=SpMV=, and \verb=BlackScholes=) can now scale to larger problems sizes.

However, we also observe that for \verb=Correlator= and \verb=K-means=, for which spilling was beneficial on one GPU in the previous section, spilling is no longer beneficial when using multiple GPUs.
For example, for \verb=K-Means=, the throughput on 1 GPU and 2 GPUs is identical for large problems.
This happens because GPUs share the PCIe bus, thus using multiple GPUs reduces the effective PCIe bandwidth per GPU.
Using multiple nodes circumvents this issue, allowing benchmarks to scale even further. 

\subsection{Multiple Nodes}
\label{sec:evaluation_dist}
Now, we present the results when using multiple nodes.
\Cref{fig:evaluation_dist_speedup} shows the throughput for up to 4 nodes with one GPU per node.
This figure looks similar to \cref{fig:evaluation_multi_speedup} since both use up to 4 GPUs, except here the GPUs are located on different nodes instead of one node.
The most notable difference is that \verb=Correlator= and \verb=K-means= can now scale to larger problem sizes that no longer fit into GPU memory since using multiple nodes means that GPUs no longer share the PCIe bus.
These benchmarks are not affected by the network overhead since InfiniBand FDR provides high bandwidth (${\sim}7$ GB/s) in the same order as PCIe 3.0 x16 (${\sim}16$ GB/s) and Lightning is able to overlap network communication with kernel execution.

Next, we scale to more than 4 GPUs.
\Cref{fig:speedup_small} shows the speedups up to 32 GPUs using 1, 2, or 4 GPUs per node.
For these experiments, we focus on weak scaling, where the problem size $n$ scales according to the number of GPUs $p$, to emphasize that our framework handles large problems far beyond the capabilities of a single GPU.
 The results show that \verb=MD5= and \verb=N-Body= scale excellently, which is expected since these benchmarks are compute-intensive and involve little data and communication.
\verb=Correlator=, \verb=K-Means=, and \verb=HotSpot= also scale near perfectly, these benchmarks do involve data but there is little communication since GPUs work on their local data.
\verb=GEMM= and \verb=SpMV= involve much communication and are more difficult to scale to more nodes.
\verb=GEMM= appears to hit the network bandwidth limit at around 16 GPUs.
\verb=Black-Scholes=' short run times make scaling difficult. 
For example, the run time on one GPU is 10.2 ms, while for a $32{\times}$ larger problem with 32 GPUs the runtime is just 10.8 ms.

\begin{figure}
  \centering
  \includegraphics[width=.65\textwidth]{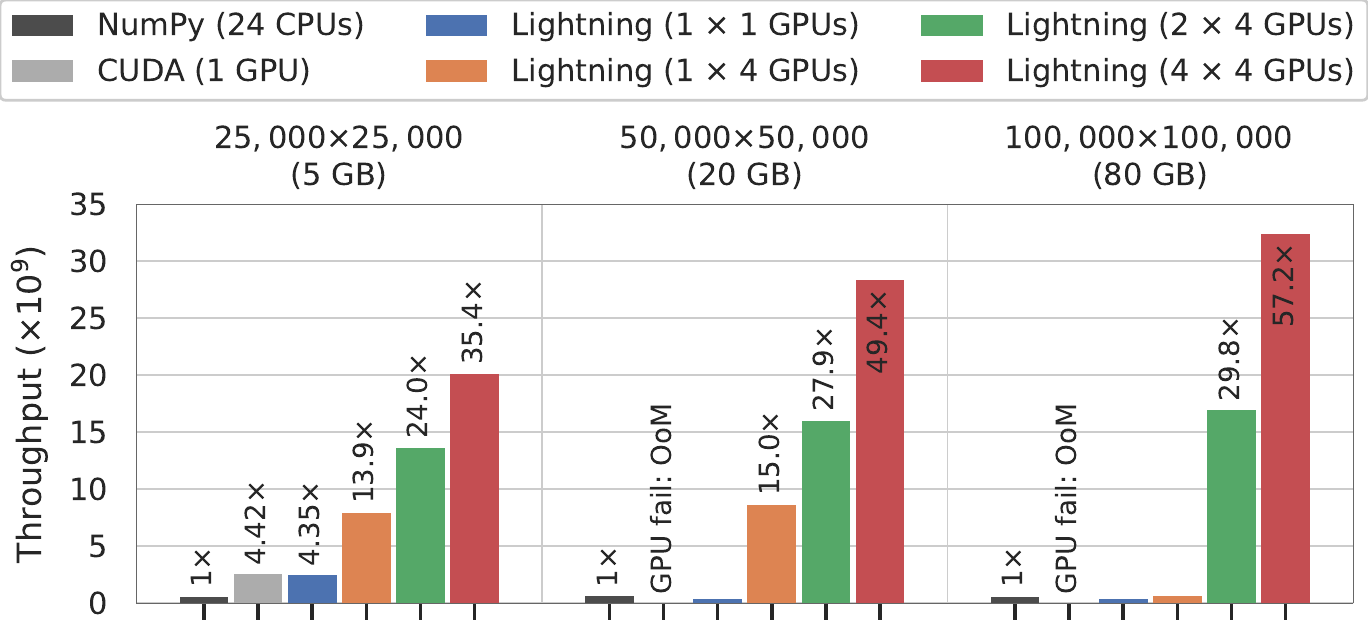}
  \caption{Performance of application for NumPy, CUDA, and Lightning on three datasets. Throughput is measured as time per iteration divided by matrix size. Notation ``\emph{$n \times m \text{ GPUs}$}'' means $n$ nodes with $m$ GPUs each. OoM is ``out of memory''.}
  \label{fig:evaluation_cgc}
\end{figure}

\subsection{Full Application}
\label{sec:evaluation_app}
In the previous sections, we considered benchmarks that are simple pipelines of one or two kernels.
To evaluate the performance of Lighting for a more complex workflow, we consider the co-clustering algorithm from CGC~\cite{cgc}: a library for geospatial cluster analysis.
Co-clustering is an iterative algorithm that clusters the rows and columns of a matrix where these dimensions correspond to space and time.
This algorithm can be used, for example, to study the impact of climate change based on the onset of spring in Europe~\cite{wu2016}.
Each iteration involves three reductions (reduction along the rows, along the columns, and along all entries), leading to a communication-intensive workload on multiple GPUs.

The original code was implemented in Python and accelerated by NumPy.
We manually reimplemented this algorithm in CUDA and tuned the resulting 10 CUDA kernels using Kernel Tuner~\cite{kerneltuner}, resulting in 635 lines of CUDA code.
Next, these kernels were adapted for use in Lightning by modifying 44 lines of code.
\Cref{fig:evaluation_cgc} shows the performance for NumPy, CUDA, and Lightning for three input matrices: 5 GB (fits into memory of 1 GPU), 20 GB (fits into 4 GPUs), and 80 GB (fits into 16 GPUs).
Performance is measured as throughput, i.e., matrix size divided by iteration time.

The results show that for the smallest matrix, the CUDA version is $4.42{\times}$ faster than the CPU version.
Lightning is $4.35{\times}$ faster, meaning an overhead of just 1.6\% over using CUDA directly.
This is anticipated since both use the same kernel code, on the same device, for the same dataset.
The plots also show that the CUDA version cannot scale to handle the larger datasets that exceed GPU memory.
For the largest matrix (80 GB), the CUDA version on one GPU fails while Lightning on 16 GPUs still works and is $57.2{\times}$ faster than NumPy on the CPU (0.31 sec versus 17.1 sec per iteration).


\section{Related Work}
\label{sec:related_work}
%
%
%

%
%

GPU programmers have a wide range of options available for creating GPU applications. 
In general, creating distributed multi-GPU applications can be achieved by
1) switching to a different programming paradigm or a combination thereof, or
2) using a system that extends the capabilities of the existing GPU programming paradigm.


\correction{
In the first category, we consider the combination of CUDA/OpenCL with, for example, MPI and OpenMP.
We also consider extensions that have been proposed to support GPUs within Big Data frameworks (e.g., 
Hadoop~\cite{zhu2014gpuinhadoop}, 
Spark~\cite{yuan2016sparkgpu}, 
Dask~\cite{daskcuda}) 
or GPU support in common HPC frameworks (e.g., 
Chapel~\cite{hayashi2019gpuiterator}, 
Charm++~\cite{vasudevan2013gcharm}, 
Legion~\cite{bauer2012legion}, 
OmpSs~\cite{bueno2012productive},
PARSeC~\cite{wu2015hierarchical},
Global Arrays~\cite{tipparaju2021ga}).
The downside of these frameworks is that GPU developers have to learn a new programming paradigm which is different from what they are used to.
In addition, while these framework give the programmer more control, they also make the programmer responsible for writing complex code to, for example, manage GPU memory, move data, split work into smaller jobs, and overlap computation and communication.
Instead, Lightning allows GPU programmers to interact with a multi-GPU cluster as if there existed a single large virtual GPU.
Programmers can create arrays and launch kernels as they are used to, while Lightning automatically distributes the work/data. 
}

There have been previous studies that also propose extensions to existing GPU programming models (CUDA and OpenCL) to facilitate the creation of distributed or multi-GPU applications.
Here, we distinguish two different approaches in the literature: A) Frameworks give explicit control over remote GPUs, and B) frameworks that implicitly distribute work across multiple GPUs. 

\subsection{Explicit control over multiple/remote GPUs}


There have been several projects that allow remote GPUs to be used as though there were local. 
Some project aim at virtualization of remote GPUs in the context of cloud computing, such as
{GridCUDA}~\cite{liang2011gridcuda}, 
{rCUDA}~\cite{duato2010rcuda}, 
{gVirtuS}~\cite{giunta2010gpgpu}, and 
{DS-CUDA}~\cite{oikawa2012dscuda}.
Strengert et al.~\cite{strengert2008cudasa} propose an interesting extension to the CUDA model that extends CUDA's three-level parallelism hierarchy (thread, block, grid) with additional levels (\emph{bus}, \emph{network}, \emph{application} levels).
For OpenCL, there have also been several attempts to provide access to remote devices, for example,
{SnuCL}/{SnuCL-D}~\cite{kim2012snucl,kim2016distributed}, 
{Distributed OpenCL} (dOpenCL)~\cite{kegel2012dopencl},
{clOpenCL}~\cite{alves2013clopencl},
{LibWater}~\cite{grasso2013libwater}, 
{HybridOpenCL}~\cite{aoki2010hybrid},
{EngineCL}~\cite{nozal2020enginecl}, 
and {dOCAL}~\cite{rasch2020docal}.

However, all the above solutions purposely do not offer any abstraction over the direct CUDA/OpenCL API, meaning programmers must manually divide the work, partition the data, and perform data transfers between GPUs.
Lightning, on the other hand, allows programmers to use a cluster of GPUs in a way that resembles single GPU programming.

\subsection{Implicitly scaling to multiple/remote GPUs}

There have been a few previous works that attempt to abstract multiple physical GPUs into a single virtual GPU.
Kim et al.~\cite{kim2011achieving} present a framework that offers multiple GPUs in one node as a single virtual OpenCL device.
Launching a kernel onto this virtual device will automatically distribute the workload and transfer the data between host and GPU memory.
There are four key aspects in which this work differs from Lighting:
1) only a single node is supported;
2) each array is entirely allocated in host memory which limits scalability;
3) workload is automatically partitioned using heuristics which forbids performance tuning and takes away control from the programmer;
4) access patterns are determined by using runtime sampling which has a runtime overhead, and can lead to misclassification, whereas Lightning's data annotations have no runtime overhead and ask programmers to consider the access pattern of their kernels.

\emph{DistCL}~\cite{diop2013distcl} is another framework that offers multiple GPUs as a single virtual OpenCL device, while also supporting clusters of GPUs.
There are three key differences between DistCL and Lightning:
1) each array is entirely allocated in the GPU memory of each device which limits scalability;
2) workloads are always partitioned along the most significant dimension, whereas Lightning allows custom workload distribution policies;
3) DistCL requires the programmer to write special \emph{meta-functions} that indicate intervals accessed by each kernel, whereas Lightning's data annotations present a more intuitive declarative approach.

\emph{MAPS-Multi}~\cite{ben-nun2015memory} is most closely related to Lightning.
MAPS-Multi is a multi-GPU programming system that facilitates workload distribution across multiple GPUs in a single node using a set of predefined data access patterns. 
Lightning is more flexible, allowing any linear data access pattern.
MAPS-Multi requires substantial modifications to CUDA kernel code, for example, for-loops are replaced with custom macros and data needs to be explicitly committed to memory.
Lightning, on the other hand, allows for existing CUDA kernels to be reused.
MAPS-Multi makes programmers responsible for data synchronization, whereas Lightning automatically takes care of this and overlaps inter-/intra-node communication and GPU computations.
Lightning also supports distributed computing over GPUs in multiple nodes, while MAPS-Multi does not.

\section{Conclusions \& Future Work}
\label{sec:conclusion}

In this work, we presented Lighting: a framework that enables GPU kernels to run on any amount of data and run on any number of GPUs, even across different nodes.
Our solution offers abstractions for \emph{distributed kernel launches} and \emph{distributed arrays} that enable transparent distribution of work and data across multiple GPUs.
Data annotations allow the framework to infer data requirements and data dependencies.
Lightning obtains excellent performance through asynchronous processing by overlapping plan construction, scheduling, data movement, and kernel execution.
Lightning is available online as open source software~\cite{lightning}\footnote{\url{https://github.com/lightning-project}}.

Evaluation shows great results. 
We observe that spilling to host memory allows data-intensive applications to work on massive data sets.
Experiments on four GPUs on a single node show excellent speedups, except spilling becomes less beneficial since GPUs on one node share PCIe bandwidth, which can be overcome by using multiple nodes.
Spilling to disk appears to be not beneficial due to limited disk bandwidth, it is possible that faster non-volatile memory (NVM) could provide a solution here.
The geospatial clustering application shows that our framework can handle large datasets, for example, processing 80 GB with 16 GPUs is $57.2{\times}$ faster than the CPU-version. 
Processing this dataset using one GPU would be impractical in terms of memory and processing power.

There are several avenues for future work.
Lightning's model is language-agnostic and support for other languages besides CUDA is in progress (e.g., OpenCL).
Additionally, Lightning currently requires manual selection of work/data distributions. 
We are working on assistance in this selection (e.g., via profiling) or even automatic selection (i.e., more intelligent planner).
There are also various interesting future topics that we did not touch upon, such as load-balancing, heterogeneous platforms, and fault-tolerance.

\bibliographystyle{plainurl}
\bibliography{lib_full}

\end{document}